 \documentclass[10pt, draft, showpacs, amsmath,amsfonts, amssymb,eqsecnum,prd, twocolumn]{revtex4}
\usepackage{bm}
\usepackage{latexsym}
\usepackage{dcolumn}
\usepackage{amsfonts,amssymb}
\usepackage{graphicx,epsfig}
\usepackage{amsmath}
\begin{document}

\def\labe{\marginpar}
\def\MA{\marginpar}
\def\a{\alpha}
\def\b{\beta}
\def\g{\gamma}
\def\d{\delta}
\def\ep{\varepsilon}
\def\vt{\vartheta}
\def\vp{\varphi}
\def\A{{\cal{A}}}
\def\ka{\kappa}
\def\pa{\partial}
\def\a{\alpha}
\def\b{\beta}
\def\m{\mu}
\def\n{\nu}
\def\g{\gamma}
\def\d{\delta}
\def\g{\gamma}
\def\la{\lambda}
\def\t{\theta}
\def\s{\sigma}
\def\r{\rho}
\def\ta{{\tilde{a}}}
\def\tb{{\tilde{b}}}
\def\tg{{\tilde{c}}}
\def\tm{{\tilde{m}}}
\def\tn{{\tilde{n}}}
\def\hi{{\hat{i}}}
\def\hj{{\hat{j}}}
  \def\hk{{\hat{k}}}
  \def\hm{{\hat{m}}}
  \def\hn{{\hat{n}}}
\def\G{{\mathcal{G}}}

\def\oG{\stackrel{{\rm o}}{\Gamma}{\!}}
  \def\oT{\stackrel{{\rm o}}{T}{\!}}
\def\LG{\stackrel{{\rm *}}{\Gamma}{\!}}
  \def\brr{\begin{eqnarray}}
\def\err{\end{eqnarray}}
\def\brn{\begin{eqnarray*}}
\def\ern{\end{eqnarray*}}


\title{A generalized photon propagator}

\author{Yakov Itin}

\affiliation{Institute of Mathematics, Hebrew University of
   Jerusalem \\
   and Jerusalem College of Technology\\
  email: {\tt itin@math.huji.ac.il}}

\begin{abstract}
A covariant gauge independent derivation of the generalized dispersion relation
of electromagnetic waves in a medium with local and linear constitutive law is
presented. A generalized photon propagator is derived. For Maxwell constitutive
tensor, the standard light cone structure and the standard Feynman propagator
are reinstated.
\end{abstract}
\date{\today}
  \maketitle
  \section{Introduction}
From a formal point of view \cite{Post}, \cite{Birkbook}, the Maxwell
electrodynamics theory can be represented  by a system of two independent
equations
  \begin{equation}\label{Max}
\epsilon^{ijkl}F_{jk,l}=0\,,\qquad H^{ij}{}_{,j}=J^i\,,
  \end{equation}
where two independent antisymmetric tensors, the field strength tensor $F_{ij}$
and the excitation tensor density $H^{ij}$  are involved. The electromagnetic
current vector density is denoted by $J^i$  Here  the commas stand for ordinary
derivatives, the indices range from 0 to 3, the Levi-Civitas permutation tensor
is normalized by $\epsilon^{0123}=1$.

 For most applications it is enough to assume a local, linear, homogeneous
 constitutive relation
between the fields  $F_{ij}$  and  $H^{ij}$,
  \begin{equation}\label{rel}
H^{ij}=\frac 12 \chi^{ijkl}F_{kl}\,.
\end{equation}
By the definition, the constitutive tensor $\chi^{ijkl}$ has to respect the
symmetries of the fields $F_{ij}$  and $H^{ij}$,
\begin{equation}\label{sym}
\chi^{ijkl}=\chi^{[ij]kl}=\chi^{ij[kl]}\,.
\end{equation}
Hence  it   has, in general,  36 independent components.

The standard Maxwell electrodynamics in vacuum is
  reinstated in this formalism by a special choice of  the Maxwell-Lorentz
  constitutive tensor
  \begin{equation}\label{M-con}
  ^{\tt{(Max)}}\chi^{ijkl}= \lambda_0\sqrt{-g}(g^{ik}g^{jl}-g^{il}g^{jk})\,.
  \end{equation}
  Here $g^{ij}$ is the Lorentz metric components,
  while $\lambda_0$ is a constant with the dimension of an admittance.

In this paper, we study the formal scheme (\ref{Max}), (\ref{rel}) with a
general constitutive tensor $\chi^{ijkl}$. The physical spacetime is considered
as a bare manifold without metrics or connection. All the information about the
geometry of this space is encoded in the constitutive tensor. In other words we
are dealing with a premetric electrodynamics.

Such a construction is  applicable  for description of a rather wide range of
physics effects. As a classical field theory, the premetric electrodynamics
  involves the standard Maxwell electrodynamics in vacuum and even provide a
  possibility to describe the additional degrees of freedom (axion,
  dilaton and skewon) as the  premetric partners of photon \cite{Obukhov:2004zz},
   \cite{serbia}.
  Moreover, since the metric is a secondary quantity in this scheme, its form
  \cite{Obukhov:2000nw}, \cite{Lammerzahl:2004ww},
  \cite{Itin:2005iv}, \cite{hehl},
and the signature \cite{Itin:2004qr} are derived from the properties of
  the constitutive tensor.
The nonminimal coupling of the electromagnetic field to the torsion yields the
birefringence of vacuum \cite{Solanki:2004az}, \cite{Preuss:2004pp}. This
effect find its natural description in the premetric scheme,
\cite{Rubilar:2003uf}, \cite{Itin:2003hr}.

Another interesting area of application is the models with violation of Lorenz
invariance. In particular the Carroll-Field-Jackiw modification of the Maxwell
electrodynamics \cite{Carroll:1989vb}, see also \cite{Kostelecky:2002hh}, is
embedded in the  premetric scheme. The wave propagation in this model requires,
however, to go beyond the geometrical optics approximation \cite{Itin:2004za}.
This problem will be considered in a contributed publication.

The  mathematical methods similar to used here was shown to be useful in ray
optics applications to GR \cite{Perlick} and in quantum plasmadynamics
\cite{Melrose}.

In the present letter, we give a covariant gauge independent derivation of the
generalized dispersion relation for the premetric electrodynamics. Moreover, we
derive a generalized Green function in the momentum representation -- a
generalized photon propagator.
\section{Dispersion relation}
 To study the wave propagation in the premetric electrodynamics model, we solve
 the first equation of (\ref{Max}) in term of potentials
 $F_{ij}=(1/2)(A_{i,j}-A_{j,i})$. Substituting it into (\ref{rel}) and (\ref{Max})
 and the current $J^i$  to be equal to zero we derive
\begin{equation}\label{equat}
\chi^{ijkl}A_{k,lj}=0\,.
\end{equation}
To study the wave-type solutions of this equation we consider an anzatz
 \begin{equation}\label{sol}
A_{ij}(x)=a_{i}e^{i\varphi}\,,
\end{equation}
where $\varphi=\varphi(x^i)$ while $a_i$ is a constant covector. Such solutions
always exist on sufficiently small neighborhoods even on bare manifold
\cite{Perlick}.  Denote the wave covector as $q_i=\varphi_{,i}$.

In the geometrical optics approximation, the changes of the media parameters
are neglected relative to the changes of the wave characteristics. Consequently
we come to an algebraic system
  \begin{equation}\label{main}
  M^{ik}a_k=0\,,\quad {\rm where} \quad M^{ik}=\frac 12 \chi^{ijkl}q_lq_j\,.
  \end{equation}
Due to the symmetries of the constitutive tensor (\ref{sym}) the matrix of the
system satisfies
   \begin{equation}\label{charge-gauge}
  M^{ik}q_k=0\,,\qquad M^{ik}q_i=0\,.
  \end{equation}
The first relation of (\ref{charge-gauge}) means the {\it gauge freedom} of the
vector potential while the second relation  is interpreted as a {\it charge
conservation condition}. Due to (\ref{charge-gauge}), the rows (and the
columns) of the matrix $M^{ij}$ are linearly dependent, so its determinant is
equal to zero. Moreover, the gauge relation (\ref{charge-gauge}) can be
interpreted as a fact that
  \begin{equation}\label{formsol}
  a_k=Cq_k
  \end{equation}
  is a formal solution of (\ref{main}).
  This solution does not give a contribution to the electromagnetic
field strength so it is unphysical.

An additional physically meaningful solution has to be linear independent on
(\ref{formsol}). A linear system has two or more linear independent solutions
if and only if its rank is two (or less).
  Consequently, a generalized electrodynamics system has a physically
  meaningful solution if
 \begin{equation}\label{adj-cond}
  A_{ij}=0\,.
  \end{equation}
Here we involved the adjoint matrix $A_{ij}$ -- a matrix constructed from the
cofactors of $M^{ij}$. The components of the adjoint matrix are expressed by
the derivatives of the determinant relative to the entries of the matrix
   \begin{equation}\label{Adj1}
A_{ij}=\frac{\partial\, det(M)}{\partial \,M^{ij}}
 =\frac 1{3!}\epsilon_{ii_1i_2i_3}\epsilon_{jj_1j_2j_3}M^{i_1j_1}M^{i_2j_2}M^{i_3j_3}\,.
   \end{equation}
 Since the adjoint matrix has, in general,  16 independent components it seems that
we have to require 16  independent conditions. The following algebraic  fact
shows that the situation is rather simpler.

{\bf  Proposition:} {\it If a square $n\times n$ matrix $M^{ij}$
  satisfies the relations
   \begin{equation}\label{matr-eq}
M^{ij}q_i=0\,,\qquad M^{ij}q_j=0\,
  \end{equation}
  for some nonzero vector $q_i$,
  its adjoint matrix $A_{ij}$ is represented by}
   \begin{equation}\label{Adj2}
A_{ij}=\lambda(q)q_iq_j\,.
   \end{equation}
   For a formal proof of this fact, see \cite{itin}.
Consequently, instead of (\ref{adj-cond}), we have only one  condition
\begin{equation}\label{gen-dis}
\lambda(q)=0\,.
\end{equation}
This condition is necessary to have physically meaningful solutions of the
generalized wave equation, so it is  a generalized dispersion relation.

The problem now is to derive from (\ref{Adj2}) the explicit expression for the
function $\lambda(q)$. It is provided \cite{itin} by using the fact that the
functions involved in (\ref{Adj2}) are homogeneous polynomials. In fact,
$A_{ij}$ is of the sixth order in the wave covector $q^i$, while $\lambda(q)$
is of the fourth order. Applying twice the derivatives with respect to the
components of the covector $q^i$ and using Euler's rule for the homogeneous
functions, we obtain
 \begin{equation}\label{calc5}
\lambda(q)=\frac 1{72}\, \frac {\partial^2 A_{ij}}{\partial q_i\partial q_j}\,.
   \end{equation}

 In  term of the matrix $M^{ij}$, the function $\lambda(q)$ is rewritten as
 \begin{eqnarray}\label{calc6}
 \lambda(q)&=&\frac 1{144} \epsilon_{ii_1i_2i_3}\epsilon_{jj_1j_2j_3}\Big(\frac {\partial^2
M^{i_1j_1}}{\partial q_i\partial q_j}M^{i_2j_2}+\nonumber\\
  &&\qquad \qquad 2\frac {\partial
M^{i_1j_1}}{\partial q_i}\frac {\partial M^{i_2j_2}}{\partial
q_j}\Big)M^{i_3j_3}\,.
\end {eqnarray}
This expression may be useful for actual calculations of the dispersion
relation for different media.

In order to have an explicit expression of the function $\lambda$ in term of
the constitutive tensor we have to calculate the corresponding derivatives. The
resulting dispersion relation is
\begin{eqnarray}\label{calc9}
&&\epsilon_{ii_1i_2i_3}\epsilon_{jj_1j_2j_3}
  \Big(\chi^{i_1(ij)j_1}\chi^{i_2abj_2}+\nonumber\\
  &&\quad
  4\chi^{i_1(ia)j_1}\chi^{i_2(jb)j_2}\Big)\chi^{i_3cdj_3}q_aq_bq_cq_d=0\,.
\end {eqnarray}
This equation is completely equivalent to the recently proposed \cite{Birkbook}
covariant dispersion relation
\begin{equation}\label{cov-dis2}
\epsilon_{ii_1i_2i_3}\epsilon_{jj_1j_2j_3} \chi^{ii_1ja}\chi^{bi_2j_1c}
 \chi^{di_3j_2j_3}q_aq_bq_cq_d=0\,.
   \end{equation}
 Indeed, in the special coordinate basis with $q_i=(q,0,0,0)$,
 both equations yield the same non-covariant expression. Also the direct proof
of the equivalence of two forms is acceptable \cite{Ob}.

The function $\lambda(q)$ is a fourth order polynomial. When it is separated to
a product of two non-positive defined quadratic factors   birefringence effect
emerges. This effect is well known from the classical optics. However, in the
premetric approach two light cones explicitly represent violation of Lorentz
invariance. The non-birefringence condition can be given \cite{Itin:2005iv} in
a rather simple covariant form: For an arbitrary covector $q$,
\begin{equation}\label{non-bir}
\lambda(q)\ge 0\,
 \end{equation}
has to be satisfied. For a component-wise representation of this condition, see
also \cite{Lammerzahl:2004ww}.

 Observe an important  special case. When  the skewon part absents, the
constitutive tensor respects the symmetries
\begin{equation}\label{sym-add}
\chi^{ijkl}=\chi^{klij}\,.
 \end{equation}
In this case, two terms in (\ref{calc9}) are proportional one to another. Thus
two additional expressions for the restricted dispersion relation emerge
\begin{equation}\label{disp-add1}
\epsilon_{ii_1i_2i_3}\epsilon_{jj_1j_2j_3}
\chi^{i_1(ij)j_1}\chi^{i_2abj_2}\chi^{i_3cdj_3}q_aq_bq_cq_d=0\,,
 \end{equation}
 and
\begin{equation}\label{disp-add2}
\epsilon_{ii_1i_2i_3}\epsilon_{jj_1j_2j_3}
\chi^{i_1(ia)j_1}\chi^{i_2(jb)j_2}\chi^{i_3cdj_3}q_aq_bq_cq_d=0\,.
\end{equation}

 For the Maxwell constitutive tensor (\ref{M-con}),
the matrix $M^{ij}$ takes the form
\begin{equation}\label{aaa}
^{\tt{(Max)}}M^{ij}=\lambda_0\sqrt{-g}\left(g^{ij}q^2-q^iq^j\right)\,.
 \end{equation}
 The corresponding adjoint matrix is
\begin{equation}\label{bbb}
^{\tt{(Max)}}A_{ij}=-\left(\lambda_0\sqrt{-g}\right)^3q^4q_iq_j\,.
 \end{equation}
 Consequently, in this special case, the dispersion relation takes
 its regular form $q^2=0$.
\section{Photon propagator}
Let us return to the full inhomogeneous Maxwell equation with a non-zero
current. In the "momentum" representation, it takes the form
 \begin{equation}\label{prop1}
 M^{ik}a_k=j^i\,.
 \end{equation}
 Observe that the charge conservation law is expressed
 now as
\begin{equation}\label{prop1x}
 j^iq_i=0\,.
 \end{equation}
It is useful to have a formal solution of the equation (\ref{prop1}) for a an
arbitrary given current $j_k$. Such a solution  is usually given by the Green
function or photon propagator, $D_{ij}(q)$. This tensor is defined in such a
way that the covector
\begin{equation}\label{prop2}
 a_k=-D_{ki}j^i\,
 \end{equation}
is a formal solution of (\ref{prop1}). Note that, due to the gauge invariant
and charge conservation, the propagator, $D_{ij}(q)$,  is defined only up to
addition of terms proportional to the wave covector $q_i$,
\begin{equation}\label{prop3}
 D_{ij}\to D_{ij}+\phi_iq_j+\psi_jq_i\,.
 \end{equation}
 Here the components of the covectors $\phi_i$ and $\psi_i$ are arbitrary functions
 of the wave covector.
In the standard electrodynamics, an expressions for this quantity is known as
 the Feynman propagator
\begin{equation}\label{prop3-feynman}
 D_{ij}=-\frac {g_{ij}}{\lambda_0 \,q^2\sqrt{-g}}\,,
 \end{equation}
 Note that it is  expressed by a symmetric matrix.
 Thus also the covectors $\phi_i$ and $\psi_i$ are usually taken to be equal
 one to another.
 In our general setting, $D_{ij}$ can be asymmetric. Consequently it useful to
 preserve two arbitrary covectors in (\ref{prop3}).

  Substituting (\ref{prop2}) into
(\ref{prop1}) we get
\begin{equation}\label{prop4}
\left( M^{ik}D_{km}-\d_m^i\right)j^m=0\,,
 \end{equation}
Note that the matrix $M^{ik}$ is singular, so the propagator  cannot be taken
to be proportional to the inverse of $M^{ik}$. In the standard case, it is not
a problem. Indeed, with the matrix $M^{ij}$ given by (\ref{aaa}) and with the
Feynman propagator (\ref{prop3-feynman}) we have
\begin{equation}\label{prop4x}
M^{ik}D_{km}=-\d^i_m+\lambda_0\frac {q^iq_m}{q^2}\sqrt{-g}\,.
\end{equation}
When this expression is  substituted in (\ref{prop4}) the second term
disappears due to the charge conservation equation (\ref{prop1x}) and the
equation is satisfied.

Our task is to derive an expression for the photon propagator in a general case
when the metric tensor is not acceptable.
 To deal with the singular matrix $M^{ij}$, we consider the tensor density
\begin{equation}\label{prop5}
 B_{ijkl}=\frac{\partial A_{ij}}{\partial M^{kl}}=\frac{\partial \,{\rm det}(M)}
 {\partial M^{kl}\partial M^{ij}}\,.
 \end{equation}
 This is the, so called, second adjoint (or the second adjugate compound) of
 the matrix $M^{ij}$.
 It is obtained by removing two arbitrary rows and two arbitrary
 columns from the original matrix.
 Observe that due to its definition the second adjoint tensor respects
the symmetry
\begin{equation}\label{prop5a}
B_{ijkl}=B_{klij}
\end{equation}
It is expressed by the components of the matrix $M^{ij}$ as
\begin{equation}\label{B-gen}
B_{ijkl}=\frac 12
\,\epsilon_{iki_1i_2}\epsilon_{jlj_1j_2}M^{i_1j_1}M^{i_2j_2}\,.
\end{equation}
From this expression, we read off the additional symmetries
\begin{equation}\label{prop5x}
B_{ijkl}=-B_{kjil}=-B_{ilkj}\,.
\end{equation}

Let us derive an identity involving the second adjoint tensor. The derivative
of the generalized Laplace expansion $ A_{ij}M^{ik}=0 $ relative to the enters
of the matrix $M^{rs}$ yields
\begin{equation}\label{prop8}
B_{ijrs} M^{ik}=-A_{rj}\delta^k_s\,.
 \end{equation}

We multiply now both sides of the equation (\ref{prop1}) by the tensor
  $B_{ijrs}$ to get
\begin{equation}\label{prop9}
B_{ijrs} M^{ik}a_k=B_{ijrs}j^i\,.
 \end{equation}
 Using (\ref{prop8}) we rewrite it as
\begin{equation}\label{prop10}
A_{rj}a_s=-B_{ijrs}j^i\,.
 \end{equation}
Substituting  (\ref{Adj2}) we get
\begin{equation}\label{prop12}
\lambda q_mq_na_k=-B_{imnk}j^i\,.
 \end{equation}
We are coming once more to the same problem: How to "divide"  both sides of
this equation by the covector $q_i$ in a covariant manner? Observe that
$\lambda$ and $B_{imnk}$ are homogeneous polynomials in $q$ of the order 4.
Assuming $j^i$ to be independent on $q$, we see that $a_k$ is a homogeneous
polynomial in $q$ of the order $-2$. Note that this is in a correspondence with
the classical expressions (\ref{prop3-feynman}).

Applying twice the partial derivatives with respect to the components of the
wave covector and using  Euler's rule for the homogeneous functions  we come to
\begin{equation}\label{prop13}
a_k=-\frac 16 \frac{\partial^2}{\partial q_m\partial q_n}
\left(\frac{B_{imnk}}\lambda\right)j^i\,.
 \end{equation}

Consequently we derived an expression for the generalized photon propagator
\begin{equation}\label{prop13x}
D_{ij}=\frac 16 \frac{\partial^2}{\partial q_m\partial q_n}
\left(\frac{B_{mijn}}\lambda\right)\,.
\end{equation}
Using the homogeneity of the polynomials involved here we get certain
equivalent expressions
\begin{equation}\label{prop13xx}
D_{ij}=\frac 1{42\lambda} \frac{\partial^2B_{mijn}}{\partial q_m\partial q_n}
=\frac 1{42\lambda}\frac{\partial^2}{\partial q_m\partial q_n}\left(
\frac{\partial A_{mi}}{\partial M^{jn}}\right)\,.
\end{equation}
In term of the matrix $M^{ij}$ it takes the form
\begin{equation}\label{prop14x}
D_{ij}=\frac
1{84\lambda}\epsilon_{imm_1m_2}\epsilon_{jnj_1j_2}\frac{\partial^2}{\partial
q_m\partial q_n}\left(M^{j_1m_1}M^{j_2m_2}\right)\,.
\end{equation}
And finally we derive an expression the generalized photon propagator via the
constitutive tensor
\begin{eqnarray}\label{prop14xx}
D_{ij}&=&\frac 1{84\lambda}\epsilon_{imm_1m_2}\epsilon_{jnj_1j_2}
\Big[\chi^{j_1(mn)m_1}\chi^{j_2abm_2}\nonumber\\
&&\qquad+2\chi^{j_1(ma)m_1}\chi^{j_2(nb)m_2}\Big]q_aq_b\,.
\end{eqnarray}

 For the Maxwell
constitutive tensor, the second adjoint takes the form
\begin{equation}\label{BMax}
B_{ijkl}=2\lambda_0^2gq^2\Big[(g_{ij}q_lq_k+g_{kl}q_iq_j)-
(g_{il}q_jq_k+g_{kj}q_iq_l)\Big]\,.
\end{equation}
Calculating with (\ref{prop13xx}) we come to the  standard Feynman propagator
expression.

\section*{Acknowledgment}
   I would like to thank Friedrich Hehl, Roman Jackiw and Volker Perlick
   for most fruitful comments. My deep acknowledgments to Yuri Obukhov for his
   sophisticated  calculations.


\begin{thebibliography}{99}
\bibitem{Post}  E.J. Post, {\it Formal Structure of Electromagnetics}
  (North Holland:Amsterdam, 1962, and Dover:Mineola, New York, 1997).

\bibitem{Birkbook} F.W.~Hehl and Yu.N.~Obukhov, {\it
     Foundations of Classical Electrodynamics: Charge, Flux, and
     Metric} (Birkh\"auser: Boston, MA, 2003).

\bibitem{Obukhov:2004zz}
  Y.~N.~Obukhov and F.~W.~Hehl,
  Phys.\ Rev.\  D {\bf 70}, 125015 (2004)

 \bibitem{serbia} F.W. Hehl, Y. Itin, Yu.N. Obukhov, Recent developments in
premetric classical electrodynamics, Proceedings of the 3rd Summer School in
Modern Mathematical Physics, 20-31 August 2004, Zlatibor, Serbia and Montenegro
B. Dragovich et al., eds., SFIN (Notebooks on Physical Sciences) XVIII:
Conferences, A1 (2005) 375-408 (Institute of Physics: Belgrade, 2005);
[arXiv.org/physics/0610221].




\bibitem{Obukhov:2000nw}
   Y.~N.~Obukhov, T.~Fukui and G.~F.~Rubilar,
   Phys.\ Rev.\ D {\bf 62}, 044050 (2000)

\bibitem{Lammerzahl:2004ww}
   C.~Lammerzahl and F.~W.~Hehl,
   Phys.\ Rev.\ D {\bf 70}, 105022 (2004)

   \bibitem{Itin:2005iv}
  Y.~Itin,
  Phys.\ Rev.\  D {\bf 72}, 087502 (2005)


\bibitem{hehl}F.W. Hehl, Yu.N. Obukhov, Spacetime metric from local and linear
electrodynamics: a new axiomatic scheme, Lecture Notes in Physics (Springer),
Vol.702 (J. Ehlers and C. L¨ammerzahl, eds.), pp.163–187 (2006);


\bibitem{Itin:2004qr}
  Y.~Itin and F.~W.~Hehl,
  Annals Phys.\  {\bf 312}, 60 (2004)


\bibitem{Solanki:2004az}
  S.~K.~Solanki {\it et al.},
  Phys.\ Rev.\  D {\bf 69}, 062001 (2004)

\bibitem{Preuss:2004pp}
   O.~Preuss, M.~P.~Haugan, S.~K.~Solanki and S.~Jordan,
   Phys.\ Rev.\ D {\bf 70}, 067101 (2004)

\bibitem{Rubilar:2003uf}
   G.~F.~Rubilar, Y.~N.~Obukhov and F.~W.~Hehl,
   Class.\ Quant.\ Grav.\  {\bf 20}, L185 (2003)

\bibitem{Itin:2003hr}
   Y.~Itin and F.~W.~Hehl,
   Phys.\ Rev.\ D {\bf 68}, 127701 (2003)



\bibitem{Carroll:1989vb}
   S.~M.~Carroll, G.~B.~Field and R.~Jackiw,
   Phys.\ Rev.\ D {\bf 41}, 1231 (1990).

\bibitem{Kostelecky:2002hh}
   V.~A.~Kostelecky and M.~Mewes,
   Phys.\ Rev.\ D {\bf 66}, 056005 (2002)

\bibitem{Itin:2004za}
  Y.~Itin,
  Phys.\ Rev.\  D {\bf 70}, 025012 (2004)

\bibitem{Perlick} V. Perlick, {\it Ray Optics, Fermat's Principle, and Applications to General
Relativity} (Springer, Berlin, 2000).


\bibitem{Melrose}
 D. B. Melrose, Plasma Phys. {\bf 15}, 99 (1973)


\bibitem{Ob} Y.~N.~Obukhov, Private communication.
\bibitem{itin} Y.~Itin, "On light propagation in premetric electrodynamics", in
preparation.

 \end{thebibliography}
\end{document}